\documentclass[12pt]{article}

\usepackage[english]{babel}
\usepackage[utf8]{inputenc}
\usepackage{amsmath,amssymb}
\usepackage{graphicx}
\usepackage{subfigure}
\usepackage[colorinlistoftodos]{todonotes}

\numberwithin{equation}{section}

\begin{document}

\title{Connection vs metric description for non-AdS solutions in higher spin theories}
\author{Yang Lei\footnote{yang.lei@durham.ac.uk}\  and Simon F. Ross\footnote{s.f.ross@durham.ac.uk} \\  \bigskip \\
Centre for Particle Theory, Department of Mathematical Sciences \\ Durham University \\ South Road, Durham DH1 3LE}
 \date{}
 \maketitle

\begin{abstract}
We consider recently-constructed solutions of three dimensional $SL(N,\mathbb{R}) \times SL(N,\mathbb{R})$ Chern-Simons theories with non-relativistic symmetries. Solutions of the Chern-Simons theories can generically be mapped to solutions of a gravitational theory with a higher-spin gauge symmetry. However, we will show that some of the non-relativistic solutions are not equivalent to metric solutions, as this mapping fails to be invertible. We also show that these Chern-Simons solutions always have a global $SL(N,\mathbb{R}) \times SL(N,\mathbb{R})$ symmetry. We argue that these results pose a challenge to constructing a duality relating these solutions to field theories with non-relativistic symmetries. 
\end{abstract}

\section{Introduction}

There has recently been considerable interest in higher spin gravity, particularly in the context of holography \cite{Gaberdiel:2012uj,Ammon:2012wc,Giombi:2012ms}. As in Einstein gravity, the three-dimensional case is particularly simple, and provides a useful laboratory for exploring the issues. The higher spin theory in three dimensions is simply a Chern-Simons theory: in general it is based on the infinite-dimensional $hs(\lambda) \times hs(\lambda)$ gauge group, but for integer values of $\lambda$ it reduces to the finite-dimensional  $SL(N,\mathbb{R}) \times SL(N,\mathbb{R})$ \cite{Blencowe:1988gj,Bergshoeff:1989ns,Henneaux:2010xg,Campoleoni:2010zq,Gaberdiel:2010pz}. From the Chern-Simons perspective it is evident that this theory has no local degrees of freedom. This includes the case of pure gravity for $N=2$. In this case it is well-known that the Chern-Simons theory corresponds to a first-order description of pure gravity with a negative cosmological constant, with the spacetime vielbein being obtained as $e_\mu = A_\mu - \bar A_\mu$, where $A, \bar A$ are the two $SL(2,\mathbb{R})$ Chern-Simons fields \cite{Witten:1988hc,Achucarro:1987vz}. Similarly the theory for integer $N$ corresponds to a theory of Einstein gravity coupled to massless fields of spin up to $N$, which are all constructed from the ``zuvielbein" $e_\mu = A_\mu - \bar A_\mu$, which is now an $SL(N,\mathbb{R})$ valued one-form.   

For any $N$, the solutions of the Chern-Simons theory include all the solutions of the  $SL(2,\mathbb{R}) \times SL(2,\mathbb{R})$ theory, so pure gravity solutions are also solutions of the higher spin theories. This includes asymptotically AdS$_3$ solutions, and the higher spin theory with asymptotically AdS$_3$ boundary conditions is conjectured to be dual to a 1+1 CFT with $W_N$ symmetry \cite{Gaberdiel:2010pz}. But the higher-spin theory is richer, and can include solutions which are not solutions of vacuum gravity. Our discussion will focus on the realisation of spacetimes with non-relativistic symmetries, the Lifshitz spacetime \cite{Kachru:2008yh}
\begin{equation}\label{Lifbackground}
ds^2=-r^{2z}dt^2+\frac{dr^2}{r^2}+r^2dx_i^2
\end{equation}
and the Schr\"odinger spacetime \cite{Son:2008ye,Balasubramanian:2008dm}
\begin{equation}\label{Schroedingerbackground}
ds^2=-r^{2z}dt^2-2r^2 dt d{x^-}+\frac{dr^2}{r^2}+r^2dx_i^2.
\end{equation}
These are of interest as potential holographic duals of field theories with non-relativistic symmetries. It would be particularly interesting to realise these as solutions of the higher-spin theories, as the large symmetry algebra may make it easier to explicitly identify the dual field theory. In addition, these solutions are known to have IR tidal force singularities (for $z \neq 1$ in the Lifshitz case \cite{Kachru:2008yh,Copsey:2010ya,Horowitz:2011gh} and for $1 < z < 2$ in the Schr\"odinger case \cite{Blau:2009gd}) which make their interpretation doubtful in a conventional metric theory. But in a higher-spin theory, the diffeomorphism symmetry is enhanced, and these singularities could possibly be just gauge artifacts, as in \cite{Castro:2011fm}. 

Solutions of the higher-spin theory which give metrics of this form were obtained in \cite{Gary:2012ms}, as we will review in section \ref{hsreview}. As a simple example, a $z=2$ Lifshitz solution can be obtained in $SL(3,\mathbb{R}) \times SL(3,\mathbb{R})$ Chern-Simons theory by taking the gauge connections to be 
\begin{equation} \label{LifCS} 
A = L_0 d\rho + W_2 e^{2 \rho} dt + L_1 e^\rho dx, \quad \bar A = -L_0 d\rho + W_{-2} e^{2 \rho} dt +  L_{-1} e^\rho dx, 
\end{equation}
which solves the Chern-Simons equations of motion $F = \bar F = 0$. Defining the spacetime metric as 
\begin{equation}
g_{\mu\nu} = \frac{1}{2} \mbox{tr} (e_\mu e_\nu)  
\end{equation}
reproduces the metric \eqref{Lifbackground}, with $r = e^\rho$. In the metric language, one would expect this solution to be supported by the spin-3 field 
\begin{equation}
\phi_{\mu\nu\lambda} = \frac{1}{6} \mbox{tr} (e_\mu e_\nu e_\lambda).
\end{equation}
In \cite{Gary:2014mca}, it was found that the spin-3 field has a non-zero $\phi_{txx}$ component. It is interesting to note that this breaks time reversal symmetry, so the Lifshitz solution would have to be holographically dual to some field theory with a vacuum which is not invariant under time reversal. 

But as we will discuss in section \ref{hsreview}, we can choose flat connections such that the metric takes the Lifshitz form \eqref{Lifbackground} but the spin-3 field identically vanishes. This is in conflict with the equations of motion in the metric formulation, as the Lifshitz metric is not a solution of the vacuum theory, and the stress tensor is constructed from terms quadratic and higher order in the spin-3 field $\phi_{\mu\nu\rho}$.  It also suggests that the breaking of time-reversal symmetry is not essential to the Lifshitz solutions. 

In section \ref{deg}, we will argue that the solution of this puzzle is that the relation between the Chern-Simons and metric formulations fails for the solution \eqref{LifCS}. In the pure gravity case $N=2$,  it is well-known that there are solutions of the Chern-Simons theory which do not correspond to regular solutions in the metric description: the vielbein $e = A - \bar A$ may fail to be invertible, implying that the metric is degenerate. The relation between the Chern-Simons and metric formulations for the $SL(3,\mathbb{R}) \times SL(3,\mathbb{R})$ Chern-Simons theory was studied in \cite{Campoleoni:2012hp,Fredenhagen:2014oua,Fujisawa:2012dk,Fujisawa:2013lua}. In particular, \cite{Fujisawa:2012dk,Fujisawa:2013lua} give a generalization of the non-degeneracy condition for the vielbein. We will see that this condition is not satisfied for the Chern-Simons fields \eqref{LifCS}. Thus, we do not have access to a metric-like formulation for this case. The cases which give a Schr\"odinger metric involve $N >3$, so we need to analyse the equivalence between Chern-Simons and metric formulations from first principles; we will find that the $z=2$ Schr\"odinger solutions are non-degenerate but the $1 < z < 2$ solutions are degenerate. We will also comment in passing that the realisations of AdS via non-principal embeddings \cite{Ammon:2011nk} also have a degenerate frame.

One might hope that this is basically a technical issue and that one could still use these solutions to explore non-relativistic holography in a Chern-Simons language: the connections \eqref{LifCS} are solutions of the flatness conditions, and they manifestly exhibit a non-relativistic scaling. However, as we will discuss in section \ref{isom}, the set of gauge transformations that leaves \eqref{LifCS} invariant is a global $SL(3,\mathbb{R}) \times SL(3,\mathbb{R})$ subgroup of the $SL(3,\mathbb{R}) \times SL(3,\mathbb{R})$ gauge group, just as in the AdS case.  This is because the solutions have no holonomies, so they can be related to $A = \bar A =0$ globally by a single-valued gauge transformation. As a result, the symmetry group is the same as that of  $A = \bar A =0$. This provides a general understanding of a fact which was uncovered as something of a surprise in the analysis of asymptotically Lifshitz solutions in \cite{Gary:2014mca}. 

If we could legitimately pass to a metric formulation, this could be separated into the Lifshitz isometries of the metric \eqref{Lifbackground} and some higher-spin gauge transformations, but in the Chern-Simons language there is nothing to pick out the Lifshitz subgroup of $SL(3,\mathbb{R}) \times SL(3,\mathbb{R})$ as special. Thus, purely in the Chern-Simons formulation, it is not clear how we identify these backgrounds as non-relativistic, in the sense that their field theory duals would have a non-relativistic symmetry. This is consistent with the results of \cite{Gary:2014mca}, which concluded that the dual of the Lifshitz cases is a field theory with $W_N$ symmetry, just as in the AdS case. 

For the Lifshitz case, asymptotically Lifshitz boundary conditions based on the solution  \eqref{LifCS} have been described in \cite{Afshar:2012nk,Gary:2014mca,Gutperle:2013oxa,Gutperle:2014aja,Beccaria:2015iwa}. In section \ref{asymp}, we comment on the extension of our analysis to asymptotically Lifshitz solutions, and argue that the boundary conditions of \cite{Gary:2014mca} could be re-interpreted as a novel kind of asymptotically AdS boundary conditions. Finally, we conclude in section \ref{concl} with a discussion of the significance of the degeneracy we find and prospects for further work.  

\section{Non-relativistic solutions in the higher spin theory}
\label{hsreview}

The $SL(N,\mathbb{R}) \times SL(N,\mathbb{R})$ Chern-Simons theory has action 
\begin{equation}\label{Chernsimonswithbar}
S = S_{CS}[A] - S_{CS}[\bar A], 
\end{equation}
where the Chern-Simons action is
\begin{equation} \label{Chernsimons}
S_{CS} = \frac{k}{4\pi}\int_M \text{Tr}(A \wedge dA+\frac{2}{3}A\wedge A\wedge A),
\end{equation}
where $k$ is the Chern-Simons level. The equations of motion are the flatness conditions 
\begin{equation}
F=dA+ A\wedge A=0; \qquad \bar{F}=d\bar{A}+\bar{A} \wedge \bar{A}=0.
\end{equation}

The theory is invariant under $SL(N, \mathbb R)$ gauge transformations
\begin{equation}
A \to A' = g^{-1} A g + g^{-1} dg, 
\end{equation}
and similarly for the barred sector. Since the connection is flat on-shell, it is locally gauge-equivalent to $A=0$, that is in open regions we can write $A = g^{-1} dg$ for some $g$. If the gauge field has holonomies they form an obstruction to writing $A$ as pure gauge globally. 

We will write solutions in the ``radial gauge'' , where we choose a radial coordinate $\rho$ and write
\begin{equation}
A = b^{-1} a b + b^{-1} db,  \qquad \bar{A} = b \bar{a} b^{-1} + b db^{-1}
\end{equation}
where $b = e^{\rho L_0}$, and $a$ is a one-form with no $d \rho$ component, which is furthermore independent of $\rho$, and a similar form is taken for the barred sector. 

This theory can be related to a higher spin gravitational theory by introducing the ``zuvielbein'' and spin connection 
\begin{equation}
e_\mu = \frac{l}{2} (A_\mu - \bar A_\mu), \quad \omega_\mu = \frac{1}{2} ( A_\mu + \bar A_\mu), 
\end{equation}
where we introduce an arbitrary length scale $l$ in defining the zuvielbein. The equations of motion then become in terms of these variables 
\begin{equation} \label{tfree}
de + e \wedge \omega + \omega \wedge e = 0, 
\end{equation}
\begin{equation}
d \omega + \omega \wedge \omega + \frac{1}{l^2} e \wedge e = 0. 
\end{equation}
In the $N=2$ case, writing $e_\mu = e^a_\mu t_a$, $e^a_\mu$ is a $3 \times 3$ matrix which we can interpret as the gravitational vielbein, and these are the equations of motion of pure gravity in a frame field formalism \cite{Witten:1988hc,Achucarro:1987vz}, with Newton constant $G_N = l/16k$. For $N>2$, $e_\mu$ is an $SL(N, \mathbb R)$ valued one-form, with $3(N^2-1)$ independent components, and it can be traded for a metric and higher-spin fields up to spin $N$. For example, for $N=3$ \cite{Campoleoni:2010zq}, we have a metric defined by 
\begin{equation} \label{metric}
g_{\mu\nu} = \frac{1}{2} \mbox{tr} (e_\mu e_\nu)  
\end{equation}
and the spin-3 field 
\begin{equation} \label{spin3} 
\phi_{\mu\nu\lambda} = \frac{1}{6} \mbox{tr} (e_\mu e_\nu e_\lambda).
\end{equation}
Henceforth we will take units with $l=1$. 

A simple class of solutions of this theory is constructed by taking the principal embedding $SL(2,\mathbb R) \subset SL(N,\mathbb R)$ and considering flat $SL(2,\mathbb R)$ connections, corresponding to vacuum gravity solutions. The global AdS$_3$ solution in Poincare coordinates is obtained by taking 
\begin{equation} \label{adsp}
a = L_1 dx^+, \quad \bar a = L_{-1} dx^-, 
\end{equation}
where $L_0, L_{\pm 1}$ are the usual $SL(2,\mathbb R)$ generators. Our conventions are set out in appendix \ref{conv}. In the metric description $x^\pm$ become null coordinates on the surfaces of constant $\rho$. 

We are interested in the non-AdS solutions constructed in \cite{Gary:2012ms}, in particular the Lifshitz and Schr\"odinger solutions. There it was found that one can construct a Lifshitz solution with integer $z$ by taking 
\begin{equation} \label{lif2}
a = a_1 W_+ dt + L_1 dx, \quad \bar a =  W_- dt + a_2 L_{-1} dx 
\end{equation}
where $W_{\pm}$ are required to satisfy
\begin{equation}
[W_{\pm}, L_0] = \pm z W_{\pm}, \quad [W_{\pm}, L_{\pm 1}] = 0, \quad tr(W_+ W_-) \neq 0, 
\end{equation}
and $a_1, a_2$ are normalization factors. For example, by taking $W_\pm = W_{\pm 2}$ in $SL(3,\mathbb R)$ we can realise Lifshitz with $z=2$; this produces the solution in \eqref{LifCS}. 

A Schr\"odinger solution with integer $z$ is obtained by taking 
\begin{equation} \label{intS}
a = (a_1 L_1 + a_2 W_+) dt, \quad \bar a =  W_- dt +  L_{-1} dx^- .
\end{equation}
With the same condition on $W_\pm$, and appropriate choices of $a_1, a_2$, this gives the metric \eqref{Schroedingerbackground}, with $r = e^\rho$. We will focus on the realisation of $z=2$ Schr\"odinger in $SL(3,\mathbb R)$ as an example of this class of solutions.  Schr\"odinger solutions with fractional weights are obtained by taking 
\begin{equation}
a = (a_1 W_+^{[1]} + a_2 W_+^{[2]}) dt, \quad \bar a =  W_-^{[2]} dt +  W_{-}^{[1]} dx^- ,
\end{equation}
where
\begin{equation}
[W_{\pm}^{[i]}, L_0] = \pm h^{[i]} W_{\pm}^{[i]}, \quad [W_-^{[1]}, W_-^{[2]}] = 0, \quad tr(W_+^{[i]} W_-^{[j]}) = t_i \delta_{ij}, t_i \neq 0. 
\end{equation}
We will take the case with $z= 3/2$ in $SL(4,\mathbb R)$ as an example of this class of solutions, where 
\begin{equation} \label{z3over2Schroedinger}
a = (U_3 +W_2) dt; \qquad \bar{a}= -\frac{5}{72}U_{-3}dt + \frac{5}{24} W_{-2}dx^{-}
\end{equation}
The corresponding metric is 
\begin{equation}
ds^2=\frac{5}{8}\left(-r^3dt^2-2r^2dtdx^-+\frac{dr^2}{r^2}\right)
\end{equation}
after replacing $r=e^{2\rho}$. 

In addition to these non-relativistic cases, we will also comment on the non-principal embeddings of AdS: for example, in $SL(3,\mathbb R)$ we can realize AdS by taking \cite{Ammon:2011nk}
\begin{equation} \label{nonpAdS}
a = W_2 dx^+, \quad \bar a = W_{-2} dx^-. 
\end{equation}

\subsection{A puzzle}

In the above solutions, we introduced some normalization constants to cancel trace factors to make the metric take the usual form with no additional numerical factors. These can be thought of as a suitable scaling of the boundary coordinates ($t,x$ or $t,\xi$ respectively). But we could go further: for example, in the $z=2$ Lifshitz case we could take 
\begin{equation} \label{glif}
a = a_1 W_2 dt + b_2 L_1 dx, \quad \bar a = b_1 W_{-2} dt + a_2 L_{-1} dx . 
\end{equation}
This is still a flat connection for any values of the constants. The metric is
\begin{equation} \label{glifa}
ds^2 = - a_1 b_1 e^{4 \rho} dt^2 + d\rho^2 + a_2 b_2 e^{2 \rho} dx^2. 
\end{equation}
We can re-absorb the constants here in redefinitions of the coordinates. But the change in the spin-3 field is more significant: the only non-vanishing component is 
\begin{equation} \label{glifb}
\phi_{txx} = - \frac{1}{4}(b_1 b_2^2 - a_1 a_2^2) e^{4 \rho}. 
\end{equation}
(Note that our conventions for the generators are different from \cite{Gary:2014mca}, as set out in appendix \ref{conv}.) 
In \cite{Gary:2014mca}, this term was interpreted as supporting the Lifshitz spacetime. It was also noted that it breaks time reversal symmetry. However, if we choose $b_1 b_2^2 = a_1 a_2^2$, we set the three-form field to zero.  How can we have a Lifshitz metric with no matter field to support it? Note that we can keep the metric fixed and change the value of the three-form field by varying the constants appropriately, so we expect that the metric equations of motion fail to be satisfied for generic values of the parameters; there might at best be some special choice of $a_1, a_2, b_1, b_2$ such that the resulting $\phi$ correctly sources the metric. 

\section{Degeneracy of the non-relativistic solutions} 
\label{deg} 

The puzzle noted above suggests that there is a problem in the relation between the Chern-Simons and metric descriptions in the Lifshitz solution. In this section we will see that there is indeed a problem for Lifshitz, some of the Schr\"odinger solutions, and AdS with non-principal embeddings. 

The issue is one that was already noted in the pure gravity case in \cite{Witten:1988hc}: the Chern-Simons description includes solutions, such as for example $A = \bar A$, for which the vielbein $e^a_\mu$ is degenerate, and hence not invertible. For pure gravity, such solutions are not acceptable solutions in the metric formulation. In addition, it is not possible to determine the spin connection in terms of the vielbein, because the vielbein is not invertible. It is this latter issue which will generalize to our case. Clearly the problem for the Lifshitz solutions is not that the metric is not invertible. But in the higher spin context, even when the metric is invertible the zuvielbein $e^a_\mu$ can fail to determine the connection $\omega^a_\mu$. 

In general, the issue is that to convert from a frame formulation of the equations to a second-order metric formulation, we want to solve the torsion-free condition \eqref{tfree} to determine the spin connection $\omega$ in terms of the zuvielbein $e$. The spin connection is an $SL(N,\mathbb{R})$ valued one-form, so it has $3(N^2-1)$ independent components. The equation is an $SL(N,\mathbb{R})$ valued two-form, so it also has $3(N^2-1)$ independent components. This is a linear algebraic system for the components of $\omega$, so generically it has a unique solution, and knowing $e$ is sufficient to determine $\omega$. In passing to the metric formulation, we exchange the information in $e$ for the metric and higher-spin fields, as in (\ref{metric},\ref{spin3}), and this data is then equivalent to the connections $A$, $\bar A$. 

But there can be special values of $e$ such that the solution of \eqref{tfree} is not unique.
(If we obtain $e = A - \bar A$ as the difference of two flat connections, then $\omega = A + \bar A$ is always a solution of \eqref{tfree}, so it can't happen that there's no solution.) The metric formulation, where we retain only the data in $e$, is then not equivalent to the Chern-Simons formulation. The two pictures are equivalent only when we can solve \eqref{tfree} for $\omega$ uniquely. 

In the $N=2$ case, we can solve \eqref{tfree} explicitly by multiplying it by the inverse frame field, so the uniqueness of solutions is equivalent  to the invertibility of $e^a_\mu$. For $N >2$, $e^a_\mu$ is not a square matrix, so we cannot express the problem in terms of its invertibility. In \cite{Fujisawa:2012dk,Fujisawa:2013lua}, this was addressed for $N=3$ by introducing additional auxiliary quantities $e^a_{\mu\nu}$ constructed out of $e^a_\mu$ such that the collection $e^a_\mu, e^a_{\mu\nu}$ forms a square matrix, and \eqref{tfree} was again explicitly solved using the matrix inverse. 

These additional quantities are constructed by first defining the symmetric tensor 
\begin{equation}
\hat{e}_{\mu\nu}=\frac{1}{2}\{e_\mu,e_\nu\}-\frac{2}{3}g_{\mu\nu} I_3
\end{equation}
where $I_3$ is the identity matrix, which is added to ensure traceless of $\hat{e}$ as a group element. Then we define the traceless tensor 
\begin{equation}
e_{(\mu\nu)}=\hat{e}_{\mu\nu}-\frac{1}{3}g_{\mu\nu}\hat{\rho}; \qquad \hat{\rho}=g^{\lambda \beta}\hat{e}_{\lambda\beta}
\end{equation}
There are five independent components of $e_{(\mu\nu)}$. Thus the combination $(e_\mu^a,e_{(\mu\nu)}^a)$ can be treated as a square matrix. In \cite{Fujisawa:2012dk}, it is shown that invertibility of this matrix is necessary and sufficient for $\omega$ to be uniquely determined by $e$. For the AdS realisation in \eqref{adsp}, \cite{Fujisawa:2012dk} show that this matrix is indeed invertible. 

Thus, for the $SL(3,\mathbb{R})$ cases, checking degeneracy reduces to checking the invertibility of this matrix. For the Lifshitz $z=2$ case, the matrix is not invertible, as 
\begin{equation}
e_{t\rho} = \hat{e}_{t\rho}=\frac{1}{2}\{e_t,e_\rho\} = \frac{1}{2} e^{2\rho} \{ a_1 W_2 - b_1 W_{-2}, L_0 \} = 0, 
\end{equation}
so the matrix has a row of zeros.  This explains why the metric-like fields we obtained in (\ref{glifa},\ref{glifb}) don't solve the equations of motion in the metric formulation: from the Chern-Simons point of view there's a higher-spin component in $\omega$ which is not determined by $g, \phi$ which plays a role in satisfying the flatness conditions. The general solution of the torsion-free condition \eqref{tfree} in this case is 
\begin{equation}
\omega=\frac{1}{2} (A +\bar{A} ) + \lambda_1 [-e^{\rho} L_0 dt + ( W_1 + W_{-1} ) dx+ \frac{1}{2} e^{-\rho} ( -W_2 + W_{-2} ) d\rho] + \lambda_2 W_0 dx
\end{equation}
where the $ \lambda_i $ are arbitrary constants parametrising the non-uniqueness of the solution.  

For the $z=2$ Schr\"odinger solution \eqref{intS}, by contrast, the matrix is invertible, so the Chern-Simons and metric formulations are equivalent. The explicit calculation is given in appendix \ref{shs}; the determinant is 
\begin{equation}
\operatorname{det} (e_{\mu}^a, e_{(\mu\nu)}^a)=-\frac{1}{32} e^{10\rho}
\end{equation}
which is non-zero for finite $\rho$. We can also check that the equations of motion in the metric formulation are satisfied by the $z=2$ Schr\"odinger fields $g,\phi$; this is discussed in appendix \ref{shs2}.

For the AdS solution in the non-principal embedding \eqref{nonpAdS}, the matrix is again not invertible. It is not hard to show $e_{++} = e_{--} = 0$. Therefore, we again have zero rows leading to vanishing determinant. The general solution for the connection $\omega$ in this case is 
\begin{equation}
\omega= \frac{1}{2}(A+\bar{A})+W_0 \Theta
\end{equation}
where $\Theta$ is an undetermined one-form. Thus, this solution does not have a metric formulation in the same metric theory as the principal embedding. However, it was argued in \cite{Campoleoni:2011hg} that for this non-principal embedding, we should consider a different metric formulation, based on interpreting the non-principal $SL(2,\mathbb{R})$ as the diffeomorphism symmetry, and decomposing the Chern-Simons field in irreducible representations of this symmetry. In this decomposition, the Chern-Simons field involves fields of lower spin (spin 1 and spin $3/2$), which will be described by first-order actions also in the metric formulation, so the map from Chern-Simons fields to this other metric formulation may not be degenerate.\footnote{We thank Andrea Campoleoni for discussion on this point.} 

Finally, we would like to consider the non-integer Schr\"odinger solutions. To do so we need to go to $N >3$, so we cannot use the description from \cite{Fujisawa:2012dk}. But for a given $e$, it is a simple linear algebra problem to check if \eqref{tfree} has a unique solution for $\omega$ or not. In the case of the $z = \frac{3}{2}$ Schr\"odinger solution in \eqref{z3over2Schroedinger}, we find that it does not have a unique solution. The general solution for the connection $\omega$ in this case is 
\begin{equation}
\omega= \frac{1}{2}(A+\bar{A})+ \hat{\omega},
\end{equation}
where the extra term $\hat{\omega}$ written in components is 
\begin{eqnarray} 
\hat{\omega}_t &=& -\frac{25}{768} \lambda_1 e^{4\rho} W_{-2} + \lambda_2 L_0- \frac{5}{8} \lambda_1 U_3+ \frac{10}{3} \lambda_2 U_0- \frac{25}{144} \lambda_3 U_{-2}- \frac{25}{432} \lambda_1 U_{-3} \\ \nonumber
\hat{\omega}_x &=& \frac{5}{32} \lambda_1 e^{3\rho}- \frac{25}{64} \lambda_1 e^{3\rho} U_{-1} \\ \nonumber
\hat{\omega}_\rho &=& - \frac{5}{8} \lambda_1 e^{\rho} W_1+ \lambda_3 L_1 + \lambda_1 L_0 + \frac{5}{3} \lambda_3 U_1 +\frac{10}{3} \lambda_1 U_0
\end{eqnarray} 
The constants $\lambda_i$ again parametrise the non-uniqueness of the solution. It is interesting to know whether second order equation of motion can be solved at perturbative order of deformation of AdS, like that we do for $z=2$ Schr\"odinger solution in Appendix \ref{shs2}. Lagrangian with the lowest order of spin-3, spin-4 fields were worked out recently \cite{Campoleoni:2014tfa}. Due to its complication, we would leave this for future work.

\section{Symmetries of the Chern-Simons solutions}
\label{isom}

In the previous section, we found that the Lifshitz solution \eqref{lif2} and the fractional $z$ Schrodinger solution  \eqref{z3over2Schroedinger} do not have a metric formulation, as the connection $\omega$ is not determined uniquely by $e$. Can we formulate a duality relating them to non-relativistic theories directly in the Chern-Simons formulation? In this section we will argue that this is challenging because the Chern-Simons formulation does not associate a distinguished set of non-relativistic symmetries with these backgrounds. 

Originally, the Lifshitz and Schr\"odinger metrics \eqref{Lifbackground} and \eqref{Schroedingerbackground} were constructed to have the corresponding symmetries as isometry groups. In the higher-spin context, these diffeomorphism isometries are supplemented by some higher-spin gauge transformations that also leave the background invariant, but one could argue that in the metric formulation we can draw a distinction between diffeomorphisms and the higher-spin gauge transformations and still regard the backgrounds as having a non-relativistic symmetry. But in the Chern-Simons formulation, it is not clear how to make such a distinction. All of the symmetries are simply gauge transformations that leave the given flat connection unchanged. 

In the discussion of asymptotically Lifshitz solutions in \cite{Gary:2014mca}, it was found that the higher-spin gauge transformations extend the Lifshitz symmetry of \eqref{Lifbackground} to a global $SL(3,\mathbb{R}) \times SL(3,\mathbb{R})$ symmetry group. In fact, there is a simple argument to see that the same happens in all cases. The symmetries are the gauge transformations $\epsilon$ such that 
\begin{equation} \label{inv}
\delta_\epsilon A = d \epsilon + [A, \epsilon] =0, 
\end{equation}
and similarly in the barred sector. The Lifshitz and Schr\"odinger metrics \eqref{Lifbackground} and \eqref{Schroedingerbackground} are analogous to AdS in Poincare coordinates, so the boundary coordinates are non-compact, and cannot be compactified without eliminating the anisotropic scaling symmetry (with the exception of the Schr\"odinger $z=2$ case, where we can compactify $\xi$). Thus, in the Chern-Simons formulation there can be no non-trivial holonomies, as there are no non-trivial topological cycles in the spacetime to measure holonomies around. As a result, the connection is globally gauge-equivalent to zero, that is each of our solutions is of the form $A = g^{-1} dg$, $\bar A = \bar g^{-1} d \bar g$ for some globally defined group elements $g$, $\bar g$. Now if we use $A = g^{-1} dg$, and set $\epsilon = g^{-1} \epsilon' g$,  \eqref{inv} reduces to 
\begin{equation}
d \epsilon' =0
\end{equation}
which is satisfied by arbitrary constant $\epsilon'$, forming a global $SL(N,\mathbb{R})$ subgroup of the gauge group. Thus the $\epsilon$ that leave $A$ invariant will always form a global $SL(N,\mathbb{R})$ group (although for a given $A$, the gauge transformations $\epsilon = g^{-1} \epsilon' g$ are not themselves constants). Thus, the symmetry of any Chern-Simons solution with no holonomies is always $SL(N,\mathbb{R}) \times SL(N,\mathbb{R})$. 

Explicitly, for the $z=2$ Lifshitz solution, $d\epsilon'=0$ can be solved by writing 
\begin{equation}
\epsilon'= \sum_{i=-1}^{1} \epsilon^{L_i}L_i +\sum_{i=-2}^{2} \epsilon^{W_i}W_i
\end{equation}
where $ \epsilon^{L_i}$ and $\epsilon^{W_i}$ are constants. The relevant group element $g$ such that $A = g^{-1} dg$ gives the Chern-Simons field in \eqref{LifCS} is $g = e^{W_2t + L_1 x} e^{\rho L_0}$. Thus the symmetries $\epsilon=g^{-1}\epsilon' g$ are
  \begin{eqnarray}
\epsilon &=& e^{\rho}(-x \epsilon^{L_0}+ \epsilon^{L_1}+x^2 \epsilon^{L_{-1}}+ t \epsilon^{W_{-1}}-4tx \epsilon^{W_{-2}})L_1 \\ \nonumber
&+& (\epsilon^{L_0}-2x \epsilon^{L_{-1}}+4t \epsilon^{W_{-2}})L_0 + e^{-\rho} \epsilon^{L_{-1}}L_{-1} \\ \nonumber
&-& e^{2\rho}(2t \epsilon^{L_0} -4tx \epsilon^{L_{-1}}-x^2 \epsilon^{W_0} +x \epsilon^{W_1} -\epsilon^{W_2} +x^3 \epsilon^{W_{-1}}+4t^2 \epsilon^{W_{-2}}-x^4 \epsilon^{W_{-2}})W_2 \\ \nonumber
&+& e^{\rho}(-4t \epsilon^{L_{-1}} -2x \epsilon^{W_0}+ \epsilon^{W_1} +3x^2 \epsilon^{W_{-1}}-4x^3 \epsilon^{W_{-2}})W_1 \\ \nonumber
&+& (\epsilon^{W_0}-3x \epsilon^{W_{-1}}+6x^2 \epsilon^{W_{-2}}) W_0 +e^{-\rho}(\epsilon^{W_{-1}}-4x \epsilon^{W_{-2}})W_{-1}+ e^{-2\rho}\epsilon^{W_{-2}} W_{-2}
\end{eqnarray}
reproducing the result of \cite{Gary:2014mca}. If we interpreted these symmetries in terms of diffeomorphisms using  $\epsilon= -\xi^\mu A_ \mu$, as suggested in \cite{Gary:2014mca},  $\epsilon^{W_2}$, $\epsilon^{L_1}$, $\epsilon^{L_0}$ parametrize time-translation, spatial translation and Lifshitz scaling respectively, although it is not clear if this is valid given that the frame is degenerate \cite{Witten:1988hc}. 

For the AdS solutions \eqref{adsp}, \eqref{nonpAdS}, the appearance of an $SL(N,\mathbb{R}) \times SL(N,\mathbb{R})$ symmetry is expected. But for the Lifshitz and Schr\"odinger solutions it implies that we cannot identify a non-relativistic isometry group from the Chern-Simons perspective. For $z=2$ Schr\"odinger, we can pass to a metric formulation, and identify the Schr\"odinger algebra as the subgroup of this  $SL(N,\mathbb{R}) \times SL(N,\mathbb{R})$ which is realised as diffeomorphisms. But for the other cases with no metric formulation there is no clear sense in which they are non-relativistic, despite the manifest scaling properties of \eqref{LifCS}; this scaling is only one of a set of $SL(N,\mathbb{R}) \times SL(N,\mathbb{R})$ symmetries. 

A possible subtlety in this argument is that when we take a background and define asymptotic boundary conditions where the fields approach the background asymptotically, the isometries of the background may not form a subgroup of the asymptotic symmetry algebra (see \cite{Brown:1986nw} for an example of this). So the non-relativistic symmetry could potentially be picked out by a notion of asymptotically Lifshitz/Schr\"odinger boundary conditions. But a choice of boundary conditions such that the asymptotic symmetry algebra does not include the symmetries of the background is usually considered undesirable. In particular, this does not happen for the asymptotically Lifshitz solutions of \cite{Gary:2014mca}, where the full $SL(3,\mathbb{R}) \times SL(3,\mathbb{R})$  symmetry is included in the asymptotic symmetry algebra. 

\subsection{Map to AdS}

One way of thinking about this result is that since all the topologically trivial solutions are gauge-equivalent to $A=\bar A = 0$, the Lifshitz and Schr\"odinger solutions can be related to the usual AdS solution by a suitable gauge transformation; so the fact that they have the same symmetries can be seen as a reflection of their just being AdS in a different gauge. Let us give this transformation explicitly in the Lifshitz case. For the AdS solution \eqref{adsp}, $A_{AdS} = g^{-1} dg$ with $g = e^{L_1 x^+} e^{L_0 \rho}$, while for the Lifshitz solution \eqref{lif2}, $A_{Lif} = h^{-1} dh$ with $h = e^{W_2 t + L_1 x} e^{\rho L_0}$. Identifying the AdS coordinate $x^+$ with $t+x$ in the Lifshitz solution, the transformation is then 
\begin{equation} 
A_{Lif} = f^{-1} df + f^{-1} A_{AdS} f ,
\end{equation}
with 
\begin{equation} \label{lifads}
f = g^{-1} h = \left( \begin{array}{ccc}
1 & 0 & 0 \\ 
-\sqrt{2}e^{\rho}t & 1 & 0 \\ 
t(t+2)e^{2\rho} & -\sqrt{2}e^{\rho}t & 1
\end{array} \right).
\end{equation}
A similar argument in the barred sector produces
\begin{equation}
\bar{f}= \left(
\begin{array}{ccc}
1 & -\sqrt{2}e^{\rho}t & t(t+2)e^{2\rho} \\ 
0 & 1 & -\sqrt{2}e^{\rho}t \\ 
0 & 0 & 1
\end{array} 
\right).
\end{equation}
We have assumed that we work with non-compact $x$, as compactifying it breaks the scaling symmetry, but it is interesting to note that compactifying $x$ does not obstruct this relation. 

\section{Asymptotically Lifshitz solutions}
\label{asymp}

So far, we have focused on the non-relativistic backgrounds, and seen that some interesting examples fail to have a corresponding metric description. Holographically, such solutions are dual to the vacuum state in the dual field theory, and it is essential to consider solutions which asymptotically approach these backgrounds to define the holographic dictionary. Since the failure of the metric description is non-generic, one would expect that considering these more generic solutions could also offer a resolution of it. In addition, imposing a given asymptotic boundary conditions partially fixes the gauge in the asymptotic region, eliminating those gauge transformations that take us out of this choice of boundary conditions. Since the bulk theory has no local degrees of freedom, it is these gauge transformations that are broken by the choice of boundary conditions that provide the physical content of the bulk theory - the higher spin analogue of the boundary gravitons. 

In this section, we will consider spacetimes which asymptotically approach the Lifshitz background \eqref{lif2}. We will first consider the asymptotically Lifshitz boundary conditions of \cite{Gary:2014mca}, which are the most well developed, and then consider alternatives. In \cite{Gary:2014mca}, asymptotically Lifshitz solutions were defined in the radial gauge as Chern-Simons solutions with 
\begin{equation}\label{generalgaugeform}
A=b^{-1}db+b^{-1}(\hat{a}^{(0)}+a^{(0)} + a^{(1)})b, \quad \bar{A}=b^{-1}db+b^{-1}(\hat{\bar{a}}^{(0)}+\bar{a}^{(0)} + \bar a^{(1)})b,
\end{equation} 
where $b=e^{\rho L_0}$, and $\hat{a}^{(0)}$, $\hat{\bar{a}}^{(0)}$ is the background solution \eqref{lif2}. The first fluctuations $a^{(0)}, \bar a^{(0)}$ have only an $x$ component, which is determined in terms of four functions $\mathcal L(x), \bar{\mathcal L}(x), \mathcal W(x), \bar{\mathcal W}(x)$, 
\begin{equation}\label{pertur}
a^{(0)}_x=4t\mathcal{W} L_0-\mathcal{L} L_{-1}-4t^2\mathcal{W} W_2+4t\mathcal{L}W_1+\mathcal{W} W_{-2},
\end{equation}
\begin{equation}\label{perturbar}
\bar{a}^{(0)}_x=-4t\bar{\mathcal{W}} L_0-\bar{\mathcal{L}} L_{1}-4t^2\bar{\mathcal{W}} W_{-2}-4t\bar{\mathcal{L}}W_{-1}+ \bar{\mathcal{W}}  W_{2}
\end{equation}
(the constant coefficients here are different from in \cite{Gary:2014mca} because we use a different convention for the $SL(3,\mathbb{R})$ generators, as set out in appendix \ref{conv}). The second subleading terms $a^{(1)}, \bar a^{(1)}$ are general, having arbitrary $t$ and $x$ components, but are required to fall off at large $\rho$, $a^{(1)}, \bar a^{(1)} \sim o(1)$. 

In \cite{Gary:2014mca}, this definition of the asymptotic boundary condition was shown to lead to finite, conserved canonical charges (constructed from the boundary functions $\mathcal L, \bar{\mathcal L}, \mathcal W, \bar{\mathcal W}$ and the gauge transformations preserving the boundary conditions) which generate a $\mathcal W_3 \oplus \mathcal W_3$ asymptotic symmetry algebra, containing the $SL(3,\mathbb{R}) \times SL(3,\mathbb{R})$ symmetries of the background \eqref{lif2}. 
\begin{eqnarray}\label{deltaL}
\delta \mathcal{L} &=&\mathcal{L}' \epsilon_L + 2\mathcal{L} \epsilon'_{L} + 3 \mathcal{W} \epsilon'_W + 2\mathcal{W}' \epsilon_W - \frac{1}{2}\epsilon'''_L \\ \nonumber
\delta \mathcal{W} &=& 3\mathcal{W} \epsilon'_L + \mathcal{W}' \epsilon_L - \frac{1}{6}\mathcal{L}''' \epsilon_W - \frac{3}{4}\mathcal{L}'' \epsilon'_W - \frac{5}{4}\mathcal{L}'\epsilon''_W - \frac{5}{6}\mathcal{L} \epsilon'''_W \\ \label{deltaW}
& + & \frac{8}{3}\mathcal{L}^2\epsilon_W' + \frac{8}{3}\mathcal{L}\mathcal{L}' \epsilon_W + \frac{1}{24}\epsilon_W^{(5)}
\end{eqnarray}
with similar expressions for the barred sector.

Because the first subleading terms do not affect the $a_t$ component, the extended vielbein at this order is still degenerate: 
\begin{equation}
e_{t\rho} =\frac{1}{2}\{e_t,e_\rho\} \approx 0, 
\end{equation}
up to terms coming from $a^{(1)}, \bar a^{(1)}$. Thus, it would seem that there are solutions with non-zero values of the charges here where the metric formulation is still not possible. For solutions with sufficiently general $a^{(1)}, \bar a^{(1)}$, the extended vielbein may be non-degenerate in the bulk, but as these terms vanish as we approach the boundary, we would expect that the inverse vielbeins of \cite{Fujisawa:2012dk} will blow up there. Thus, the degeneracy is a real obstacle to the construction of a good metric description for this class of asymptotically Lifshitz boundary conditions. 

It was argued in \cite{Gary:2014mca} that these boundary conditions are distinct from the usual asymptotically AdS boundary conditions \cite{Campoleoni:2010zq}. Two main arguments were given: one relied on the breaking of time-reversal invariance in the Lifshitz solution, but as we have seen it is possible to take the generalised backgrounds in \eqref{glif} such that the spin-three field vanishes, eliminating the breaking of time-reversal symmetry. The other was that the asymptotically Lifshitz boundary conditions involve functions of $x$, while asymptotically AdS boundary conditions involve functions of $x^{\pm}$. This indeed shows that asymptotically Lifshitz solutions are distinct from the asymptotically AdS solutions, if we relate the two backgrounds using the gauge transformation \eqref{lifads}. 

However, given the failure of the metric description in the gauge \eqref{lif2}, we think it may be more straightforward to understand the physical significance of these boundary conditions if we apply this gauge transformation to re-express them in terms of the AdS solution \eqref{adsp}. That is, let us take the solutions (\ref{pertur},\ref{perturbar}) and apply the gauge transformation \eqref{lifads}. We then obtain a family of solutions of the form \eqref{generalgaugeform}, but where now $\hat a^{(0)}$, $\hat{\bar a}^{(0)}$ are the AdS background \eqref{adsp}, and 
\begin{eqnarray}\label{AdSfixt}
a^{(0)}_x &=&- \mathcal{L} t^2L_1- 2\mathcal{L} t L_0 -\mathcal{L} L_{-1} +\mathcal{W} t^4 W_2 +4\mathcal{W} t^3 W_1 \\ &&+ 6\mathcal{W} t^2 W_0 +4\mathcal{W} t W_{-1}+\mathcal{W} W_{-2}, \nonumber \\ \nonumber
\bar{a}^{(0)}_x &=& - \bar{\mathcal{L}} t^2L_{-1}- 2\bar{\mathcal{L}}t L_0 -\bar{\mathcal{L}} L_1 +\bar{\mathcal{W}} t^4 W_{-2} +4 \bar{\mathcal{W}} t^3 W_{-1} \\ &&+ 6\bar{\mathcal{W}} t^2 W_0 +4\bar{\mathcal{W}} t W_{1}+\bar{\mathcal{W}} W_{2}. 
\end{eqnarray} 
Thus, the asymptotic boundary conditions of \cite{Gary:2014mca} can be rewritten in a different gauge as a new kind of asymptotically AdS boundary conditions. Since in this gauge the relation to the metric formulation is possible, the physics of the boundary conditions may be clearer in this gauge. Note the asymptotic  symmetry algebra \eqref{deltaL}, \eqref{deltaW} is unaffected when we shift from Lifshitz gauge solution to AdS gauge solution.

An alternative asymptotically Lifshitz boundary condition was given in \cite{Gutperle:2013oxa}. The connection is taken to have the form 
\begin{eqnarray}
 a_t&=&W_2-2\mathcal{L} W_0+\frac{2}{3}\mathcal{L}'W_{-1}-2\mathcal{W} L_{-1}+ (\mathcal{L}^2-\frac{1}{6}\mathcal{L}'')W_{-2},\\
 a_x&=&L_1-\mathcal{L}L_{-1}+\mathcal{W}W_{-2},
 \end{eqnarray}
where $\mathcal L$ and $\mathcal W$ are now functions of both $t$ and $x$, subject to the consistency conditions 
\begin{eqnarray}
 \dot{\mathcal{L}} &=&2 \mathcal{W}', \\
 \dot{\mathcal{W}} &=&\frac{4}{3}(\mathcal{L}^2)'-\frac{1}{6}\mathcal{L}'''.
 \end{eqnarray}
Similarly, for the barred fields
 \begin{eqnarray}
 \bar{a}_t&=&W_{-2}-2\bar{\mathcal{L}} W_0-\frac{2}{3}\bar{\mathcal{L}}' W_{1}+2\bar{\mathcal{W}} L_{1}+ (\bar{\mathcal{L}}^2- \frac{1}{6}\bar{\mathcal{L}}'') W_{2},\\
 \bar{a}_x&=&L_{-1}-\bar{\mathcal{L}}L_{1}-\bar{\mathcal{W}}W_{2},
 \end{eqnarray}
 with consistency constraints
\begin{eqnarray}
 \dot{\bar{\mathcal{L}}} &=&-2 \bar{\mathcal{W}}', \\
 \dot{\bar{\mathcal{W}}} &=&-\frac{4}{3}(\bar{\mathcal{L}}^2)'+\frac{1}{6}\bar{\mathcal{L}} '''.
 \end{eqnarray}
In these asymptotic boundary conditions, the degeneracy of the generalised frame is resolved for generic $\mathcal L$, $\mathcal W$. The determinant is
 \begin{equation}
-\frac{\mathcal{W}^3}{8r^{14}}(r^2+\mathcal{L})^4 [(r^2+\mathcal{L})^3-2\mathcal{W}^2][(r^2+\mathcal{L})(r^2-\mathcal{L})^2-2\mathcal{W}^2]. 
\end{equation}
There are some specific points $r$ where the determinant vanishes. These singularities would not spoil the non-degeneracy property and can be avoided by method of fibre bundle \cite{Fujisawa:2012dk}. Since the determinant is not vanishing even at large $r$, one would expect a metric formulation is possible even in the asymptotic region. It may be interesting to explore these boundary conditions further; it was noted in \cite{Gary:2014mca} that the canonical charges in this case are finite but not conserved. 

In \cite{Gutperle:2013oxa}, there was also a further generalization to turn on some source terms, taking
 \begin{eqnarray}\nonumber
 a_t&=& \mu_2 W_2+\mu_1 L_1 -2\mathcal{L}\mu_2 W_0-(2\mathcal{W}\mu_2+ \mathcal{L} \mu_1)L_{-1}+ (\mathcal{L}^2\mu_2+\mathcal{W} \mu_1)W_{-2},\\
 a_x&=&L_1-\mathcal{L}L_{-1}+\mathcal{W}W_{-2},
 \end{eqnarray}
and barred sector
\begin{eqnarray}\nonumber
 \bar{a}_t&=& \mu_2 W_{-2}-\mu_1  L_{-1}-2\bar{\mathcal{L}}\mu_2 W_0+ (2\bar{\mathcal{W}} \mu_2+ \bar{\mathcal{L}} \mu_1) L_{1}+ (\bar{\mathcal{L}}^2\mu_2 + \bar{\mathcal{W}}\mu_1) W_{2},\\
 \bar{a}_x&=&L_{-1}-\bar{\mathcal{L}}L_{1}-\bar{\mathcal{W}}W_{2}.
 \end{eqnarray}
The presence of the sources $\mu_1$, $\mu_2$ makes the determinant of the generalized vielbein non-zero even for vanishing $\mathcal L, \mathcal W$, so this deformation away from Lifshitz resolves the degeneracy of the generalized vielbein even in the vacuum. The metric formulation is well-defined in this case since metric-like fields solve Einstein equations by the method in appendix \ref{shs2}. We leave further study of these deformations to future work. 

\section{Conclusions}
\label{concl}

We have seen that the Lifshitz and non-integer Schr\"odinger solutions of \cite{Gary:2012ms} have degenerate generalized vielbeins, so they are not equivalent to some solution in the metric formulation of the higher spin theory. We also found that in all cases the symmetries of the backgrounds in the Chern-Simons formulation are $SL(N,\mathbb{R}) \times SL(N,\mathbb{R})$, generalizing and simplifying an observation of \cite{Gary:2014mca}. These seem significant obstacles to interpreting these backgrounds as non-relativistic solutions. The Schr\"odinger solutions with integer $z$ have non-degenerate generalized vielbeins, so they remain as non-trivial examples of non-relativistic backgrounds in the higher spin context. But our results prevent us from studying several interesting questions about these backgrounds, such as identifying examples of Lifshitz field theories or addressing the physical meaning of the IR singularities in the metrics (\ref{Lifbackground}, \ref{Schroedingerbackground}). 

These problems could be moderated by considering classes of solutions which asymptotically approach these backgrounds, although one would be concerned that the problem with the vacuum solution would reappear in the asymptotic region. For the most well-developed example of asymptotically Lifshitz boundary conditions in the higher spin context \cite{Gary:2014mca}, we find that the generalized vielbein is still degenerate at first subleading order. We have proposed that these boundary conditions may be more usefully viewed instead as a novel asymptotically AdS boundary condition. In that gauge a metric formulation is available, and it would be interesting to understand the differences from the usual asymptotically AdS  boundary condition. For the boundary conditions of \cite{Gutperle:2013oxa}, the degeneracy of the generalized vielbeins  was lifted, and it appeared that an inverse could exist even in the asymptotic region. It would be interesting to understand this case further. 

The problems we have found are likely to be special to the case of three bulk dimensions, as the Chern-Simons formulation is particular to this case, and the absence of bulk degrees of freedom also obstructs obtaining richer families of solutions. It would be interesting to explore the realisation of non-relativistic backgrounds like  (\ref{Lifbackground}, \ref{Schroedingerbackground}) in higher-dimensional higher spin theories \cite{Vasiliev:1995dn,Vasiliev:1999ba,Giombi:2012ms}. 

\section*{Acknowledgements}

We are grateful for discussions with Andrea Campoleoni, Mirah Gary,  Wei Li, Jos\'e M. Mart\'in-Garc\'ia, Teake Nutma, Stefan Prohazka and Stefan Theisen, and for comments on a draft from Daniel Grumiller and Mirah Gary. SFR is supported in part by STFC under grant number ST/L000407/1. 

\appendix

\section{Conventions}
\label{conv} 

\subsection{$sl(3,R)$ Algebra}
The conventions in two cases are different. The $sl(3,R)$ generators satisfy algebra \begin{equation}
[L_n,L_m]=(n-m)L_{n+m} 
\end{equation}
\begin{equation}
[L_n,W_m]=(2n-m)W_{n+m}
\end{equation}
\begin{equation}
[W_n,W_m]=\sigma (n-m)(2n^2+2m^2-mn-8)L_{m+n}
\end{equation}
In our calculation $\sigma=-\dfrac{1}{12}$. Our generators are \begin{equation} \nonumber 
L_{-1}=\left(                
  \begin{array}{ccc}   
    0 & \sqrt{2} & 0\\  
    0 & 0 & \sqrt{2}\\ 
    0 & 0 & 0\\
  \end{array}
\right), \quad
L_{1}=\left(                
  \begin{array}{ccc}   
    0 & 0 & 0\\  
    -\sqrt{2} & 0 & 0\\ 
    0 & -\sqrt{2} & 0\\
  \end{array}
\right)      , \quad  
L_{0}=\left(                
  \begin{array}{ccc}   
    1 & 0 & 0\\  
    0 & 0 & 0\\ 
    0 & 0 & -1\\
  \end{array}
\right)
\end{equation}
\begin{equation} \nonumber 
W_{-2}=\left(                
  \begin{array}{ccc}   
    0 & 0 & 2\\  
    0 & 0 & 0\\ 
    0 & 0 & 0\\
  \end{array}
\right), \quad
W_{2}=\left(                
  \begin{array}{ccc}   
    0 & 0 & 0\\  
    0 & 0 & 0\\ 
    2 & 0 & 0\\
  \end{array}
\right)      , \quad  
W_{0}=\frac{1}{3}\left(                
  \begin{array}{ccc}   
    1 & 0 & 0\\  
    0 & -2 & 0\\ 
    0 & 0 & 1\\
  \end{array}
\right)
\end{equation}

\begin{equation} \nonumber 
W_{-1}=\frac{1}{\sqrt{2}}\left(                
  \begin{array}{ccc}   
    0 & 1 & 0\\  
    0 & 0 & -1\\ 
    0 & 0 & 0\\
  \end{array}
\right), \quad
W_{1}=\frac{1}{\sqrt{2}}\left(                
  \begin{array}{ccc}   
    0 & 0 & 0\\  
    -1 & 0 & 0\\ 
    0 & 1 & 0\\
  \end{array}
\right) 
\end{equation}

\subsection{$sl(4,R)$ algebra}
Our representation of $sl(4,R)$ algebra is slightly different from \cite{Gary:2012ms}. 
\begin{equation} \nonumber 
L_{-1}=\left(                
  \begin{array}{cccc}
    0 & \sqrt{3} & 0 & 0\\  
    0 & 0 & 2 & 0\\ 
    0 & 0 & 0 & \sqrt{3}\\
    0 & 0 & 0 & 0
  \end{array}
\right), \quad
L_{0}=\frac{1}{2}\left(                
  \begin{array}{cccc}   
    3 & 0 & 0 & 0\\  
    0 & 1 & 0 & 0\\ 
    0 & 0 & -1 & 0\\
    0 & 0 & 0 & -3 
  \end{array}
\right), \quad
L_{1}=\left(                
  \begin{array}{cccc}   
    0 & 0 & 0 & 0\\  
     -\sqrt{3} & 0 & 0 & 0\\ 
    0 & -2 & 0 & 0\\
    0 & 0 & -\sqrt{3} & 0
  \end{array}
\right)
\end{equation}
Quintet: \begin{equation}\nonumber
W_2=\left(                
  \begin{array}{cccc}  
    0 & 0 & 0 & 0\\  
    0 & 0 & 0 & 0\\ 
    2\sqrt{3} & 0 & 0 & 0\\
    0 & 2\sqrt{3} & 0 & 0
  \end{array}
\right), \quad
W_{-2}=\left(                
  \begin{array}{cccc}
    0 & 0 & 2\sqrt{3} & 0\\  
    0 & 0 & 0 & 2\sqrt{3}\\ 
    0 & 0 & 0 & 0\\
    0 & 0 & 0 & 0
  \end{array}
\right)
\end{equation}

\begin{equation}\nonumber
W_{0}=\left(                
  \begin{array}{cccc}
    1 & 0 & 0 & 0\\  
    0 & -1 & 0 & 0\\ 
    0 & 0 & -1 & 0\\
    0 & 0 & 0 & 1
  \end{array} 
\right), \quad 
W_{-1}=\left(                
  \begin{array}{cccc}
    0 & \sqrt{3} & 0 & 0\\  
    0 & 0 & 0 & 0\\ 
    0 & 0 & 0 & -\sqrt{3}\\
    0 & 0 & 0 & 0
  \end{array}
\right), \quad
W_{1}=\left(                
  \begin{array}{cccc}
    0 & 0 & 0 & 0\\  
   -\sqrt{3} & 0 & 0 & 0\\ 
    0 & 0 & 0 & 0\\
    0 & 0 & \sqrt{3} & 0
  \end{array}
\right)
\end{equation}
Septet: \begin{equation} \nonumber
U_3=\left(                
  \begin{array}{cccc}  
    0 & 0 & 0 & 0\\  
    0 & 0 & 0 & 0\\ 
    0 & 0 & 0 & 0\\
    -6 & 0 & 0 & 0
  \end{array}
\right), \quad 
U_2=\left(                
  \begin{array}{cccc}
    0 & 0 & 0 & 0\\  
    0 & 0 & 0 & 0\\ 
    \sqrt{3} & 0 & 0 & 0\\
    0 & -\sqrt{3} & 0 & 0
  \end{array}
\right), \quad 
U_1=\frac{2}{5}\left(                
  \begin{array}{cccc}
    0 & 0 & 0 & 0\\  
    -\sqrt{3} & 0 & 0 & 0\\ 
    0 & 3 & 0 & 0\\
    0 & 0 & -\sqrt{3} & 0
  \end{array}
\right)
\end{equation}

\begin{equation} \nonumber
U_{0}=\frac{1}{10}\left(                
  \begin{array}{cccc}
   3 & 0 & 0 & 0\\  
    0 & -9 & 0 & 0\\ 
    0 & 0 & 9 & 0\\
    0 & 0 & 0 & -3
  \end{array} 
\right), \quad 
U_{-1}=\frac{2}{5}\left(                
  \begin{array}{cccc}
   0 & \sqrt{3} & 0 & 0\\  
    0 & 0 & -3 & 0\\ 
    0 & 0 & 0 & \sqrt{3}\\
    0 & 0 & 0 & 0
  \end{array} 
\right)
\end{equation}

\begin{equation}\nonumber
U_{-2}=\left(                
  \begin{array}{cccc}
   0 & 0 & \sqrt{3} & 0\\  
    0 & 0 & 0 & -\sqrt{3}\\ 
    0 & 0 & 0 & 0\\
    0 & 0 & 0 & 0
  \end{array} 
\right), \quad
U_{-3}=\left(                
  \begin{array}{cccc} 
   0 & 0 & 0 & 6\\  
    0 & 0 & 0 & 0\\ 
    0 & 0 & 0 & 0\\
    0 & 0 & 0 & 0
  \end{array} 
\right)
\end{equation}

\section{Schr\"odinger higher spin calculations}
\subsection{Determinant}
\label{shs}

We consider the most general form of Schr\"odinger solution after normalization: \begin{equation} \label{Schroedingernormal1}
a_t=kW_2+c L_1; \qquad a_{x^-}=0
\end{equation}
\begin{equation}\label{Schroedingernormal2}
\bar{a}_t=\frac{1}{k}W_{-2}; \qquad \bar{a}_{x^-}=\frac{2}{c}L_{-1}
\end{equation}
Dreibein $e$ can be found to be \begin{equation}
e=L_0 d\rho + \frac{1}{2} (k e^{2\rho} W_2+ c e^{\rho} L_1- \frac{1}{k}e^{2\rho} W_{-2}) dt -\frac{1}{c} e^{\rho} L_{-1} d{x^-}
\end{equation}
The extra introduced 5 tetrads are 
\begin{eqnarray}
e_{({x^-}{x^-})} &=& \frac{1}{c^2} e^{2\rho} W_{-2}\\ 
e_{(\rho\rho)} &=& W_0- \frac{1}{3c^2} e^{2\rho} W_{-2}- \frac{k}{3c} e^{\rho} L_1 \\
e_{(t{x^-})} &=& -\frac{1}{2}e^{2\rho} W_0+ \frac{k}{6c} e^{3\rho}L_1 -\frac{1}{3c^2} e^{4\rho} W_{-2} \\
e_{(tt)} &=& \frac{c^2}{4}e^{2\rho} W_2- e^{4\rho} W_0 +\frac{1}{3c^2} e^{6\rho} W_{-2} + \frac{k}{3c} e^{5\rho} L_1 +\frac{c}{2k} e^{3\rho} L_{-1} \\
e_{(t\rho)} &=& \frac{1}{2} c e^{\rho} W_1 \\
e_{(\rho {x^-})} &=& -\frac{1}{c} e^{\rho} W_{-1}
\end{eqnarray}
We only need 5 of these tetrad since they are linearly dependent due to the traceless condition $g^{\mu\nu} e_{(\mu\nu)}=0$. In this specific case, $$
e_{({x^-}{x^-})}+ e_{(\rho\rho)}+ 2e^{-2\rho} e_{(t{x^-})}=0
$$
Therefore, we calculate the determinant of $8\times 8$ matrix with spacetime indices excluding $(\rho\rho)$. \begin{equation}
\operatorname{det} (e_{\mu}^a, e_{(\mu\nu)}^a)=-\frac{1}{32} e^{10\rho}
\end{equation}
We find this nonvanishing value is independent of the choice of $k$ and $c$. Then we should be able to map frame-like Schr\"odinger solution  (\ref{Schroedingernormal1}) (\ref{Schroedingernormal2}) to metric-like fields. 

\subsection{Einstein equation in D=3 higher spin theory}
\label{shs2}
We showed that the zuvielbein of $z=2$ Schr\"odinger solution in $SL(3,R)$ has non-vanishing determinant. One would then expect the fields constructed from it to solve the equations of motion in the metric formulation. In terms of metric-like fields $g, \phi$, Lagrangian of \eqref{Chernsimonswithbar} can be written as \cite{Campoleoni:2012hp} 
\begin{equation}
\mathcal{L} = \mathcal{L}_{ \text{E-H} }+ \mathcal{L}_F,
\end{equation}
where $\mathcal{L}_{ \text{E-H} }= R+ \dfrac{2}{l^2}$ and $\mathcal{L}_F$ contains terms depending on $\phi$ (note that we set $l=1$). $\mathcal{L}_F$ was worked out to quadratic order in $\phi$ terms in \cite{Campoleoni:2012hp}, with general expression: 
\begin{eqnarray}
\mathcal{L}_F ( \phi^2 ) &=& \phi^ {\mu \nu \rho }(\mathcal{F}_{\mu \nu \rho }-\frac{3}{2}g_{(\mu\nu}\mathcal{F}_{\rho)})+m_1\phi_{\mu\nu\rho}\phi^{\mu\nu\rho}+m_2\phi_\mu \phi^\mu \\
&+& 3R_{\rho\sigma}(k_1\phi^\rho_{\mu\nu}\phi^{\sigma \mu\nu}+k_2\phi^{\rho\sigma}_\mu \phi^\mu+ k_3 \phi^\rho \phi^\sigma) + 3R (k_4 \phi_{\mu\nu\rho}\phi^{\mu\nu\rho}+k_5 \phi_\mu \phi^\mu) \nonumber 
\end{eqnarray} 
where $\phi_\rho=\phi_{\rho \mu}^{\quad \mu}$, $\mathcal{F}_\rho=\mathcal{F}_{\rho \mu}^{\quad \mu}$  and $\mathcal{F}_{\mu \nu \rho }$ is the Fronsdal tensor defined by 
\begin{equation}
\mathcal{F}_{\mu \nu \rho }=\nabla^\sigma \nabla_\sigma \phi_{\mu\nu\rho}-\frac{3}{2}( \nabla^\sigma \nabla_{(\mu}\phi_{\nu \rho) \sigma}+ \nabla_{(\mu}\nabla^\lambda \phi_{\nu\rho) \lambda} ) +3 \nabla_{(\mu} \nabla_\nu \phi_{\rho )}
\end{equation}
The mass coefficients $m_i$ are determined by requiring invariance under gauge transformations, which gives
\begin{equation}
m_1=6 ( k_1+3 k_4 - 1); \qquad m_2=6 (k_2 + k_3 + 3 k_5 + \frac{9}{4})
\end{equation} 
Different $k_i$s may parametrize the same theory if one performs a redefinition of metric and spin-3 fields. For convenience, let's take those values of $k_i$ in \cite{Campoleoni:2012hp},

\begin{equation}
k_1=\frac{3}{2}; \quad k_2=0; \quad k_3=-\frac{3}{4}; \quad k_4=-\frac{1}{2}; \quad k_5=0
\end{equation}
The unique choice of $k_i$ were determined by requiring asymptotically AdS solution solving Einstein equation. 
\begin{equation}\label{spin3einstein}
R_{\mu \nu} - \frac{1}{2} g_{\mu \nu} R - g_{\mu \nu}=- \frac{1}{ \sqrt{-g} } \frac{ \delta (\sqrt{-g} \mathcal{L}_{F} )}{\delta g^{\mu \nu}}
\end{equation}
The right-hand side of the equation is too complicated to display. Exact expression is accessible in \cite{Pfenninger}. We perform the calculation by the help of \emph{xAct} package \cite{DBLP:journals/corr/Nutma13,xact}.

Schr\"odinger spacetime is not asymptotically AdS. However, one can consider it as perturbative deformation of AdS \cite{Guica:2010sw}. The zuveilbein to our interest would be 
\begin{equation}\label{GeneralSchreodinger1}
a_t =  L_1+ \sigma W_2; \qquad a_{x^-} = 0
\end{equation}
\begin{equation}\label{GeneralSchreodinger2}
\bar{a}_t = \sigma W_{-2}; \qquad \bar{a}_{x^-} = 2 L_{-1}
\end{equation}
which corresponds to metric
\begin{equation}\label{deformSchroedinger}
ds^2 = -\sigma^2 r^{4} dt^2 + \frac{dr^2}{r^2} + 2 r^2 dt dx^-
\end{equation}
and spin-3 field 
\begin{equation}\label{Schrhigherspin}
\phi_{t--} = \frac{\sigma}{3}r^4 ; \qquad \phi_{ttt} = -\frac{\sigma}{4} r^4
\end{equation}
$\sigma$ measures deformation from pure AdS in lightcone frame. Apparently metric fields would solve Einstein equation if $\sigma=0$. 

After substituting \eqref{deformSchroedinger} and \eqref{Schrhigherspin} into \eqref{spin3einstein}, one can find the equation holds at the lowest order of $\sigma$. Similarly, one can also check the equation of motion about $\phi_{\mu \nu \rho}$ \cite{Campoleoni:2012hp} can be solved at the same order of $\sigma$.

\bibliographystyle{JHEP}
\bibliography{higherspin}

\end{document}